\documentclass[preprint,12pt,times,number]{elsarticle}

\usepackage{amsmath}
\usepackage{amssymb}
\usepackage{graphicx}
\usepackage{booktabs}
\usepackage{siunitx}
\usepackage{placeins}
\usepackage{float}
\graphicspath{{./}}

\journal{Physics Letters B}

\newcommand{\nzero}{n_0}
\newcommand{\epszero}{\varepsilon_0}
\newcommand{\cs}{c_s}
\newcommand{\cssnm}{c_{s,\SNM}}
\newcommand{\csbeta}{c_{s,\betaeq}}
\newcommand{\dd}{\mathrm{d}}
\newcommand{\SNM}{\mathrm{SNM}}
\newcommand{\betaeq}{\beta}

\begin{document}

\begin{frontmatter}

\title{Trace anomaly and isospin splitting in inverse-mapped relativistic mean-field theory}

\author[addr1,addr2]{Wen-Jie Xie\corref{cor1}}
\ead{xiewenjie@ycu.edu.cn}
\author[addr1,addr2]{Jun-Hua Guo}
\cortext[cor1]{Corresponding author}
\address[addr1]{Department of Physics, Yuncheng University, Yuncheng 044000, China}
\address[addr2]{Shanxi Province Intelligent Optoelectronic Sensing Application Technology Innovation Center, Yuncheng University, Yuncheng 044000, China}

\begin{abstract}
The dimensionless trace measure \(\Delta=1/3-P/\varepsilon\) and the sound
speed \(\cs^2=\dd P/\dd\varepsilon\) provide a compact thermodynamic language
for comparing heavy-ion collision constraints on symmetric nuclear matter with
multimessenger constraints on beta-equilibrated neutron-star matter.  We use
a uniform-matter inverse-mapped relativistic mean-field ensemble to ask whether
the nonconformal behavior inferred from these two environments can be converted
from an EOS-level statement into a microscopic channel diagnostic.  The full
posterior combines chiral
effective field theory, heavy-ion flow, and neutron-star information while
sampling the density dependence of the isoscalar-scalar, isoscalar-vector,
and isovector-vector responses simultaneously.  Its symmetric-matter trace
anomaly is consistent with recent flow-based extractions, while
beta-equilibrated matter approaches the neutron-star trace bands more slowly.
The difference acts as an isospin diagnostic within the present inverse-mapped
ensemble: the splitting
\(\Delta_{\SNM}-\Delta_{\betaeq}\) is most strongly associated with the
density derivative of the isovector coupling, with bootstrap-stable Spearman
coefficients \(r_s\simeq0.91--0.92\) at \(2--3\nzero\).  Its correlation with
the beta-equilibrium proton fraction is much weaker.  The sound-speed
splitting changes sign near \(3.38\nzero\), and the derivative contribution
\(-\dd\Delta/\dd\ln\varepsilon\) becomes associated with both scalar-vector
and isovector responses above \(4\nzero\).  Data-combination tests show that
this isovector separation appears only when laboratory and astrophysical
projections are combined.  The ensemble is used as a uniform-matter inverse
map rather than as a final finite-nucleus-calibrated energy density
functional.  A comparison with fixed density-dependent RMF baselines shows
partial overlap near \(2\nzero\), but the inverse posterior favors a more
persistent positive SNM--beta trace splitting at higher density.  Trace
constraints therefore do not by themselves identify the
microscopic composition; combined with an inverse microscopic map, however,
they identify \(\Delta_{\SNM}-\Delta_{\betaeq}\) as a new thermodynamic probe
of the high-density symmetry sector, constraining how the isoscalar
nonconformal trajectory seen in heavy-ion collisions is transported to
neutron-star matter.
\end{abstract}


\end{frontmatter}

\section{Introduction}

The equation of state (EOS) of cold dense matter is now constrained from two
directions.  Heavy-ion collision (HIC) measurements probe the pressure of
nearly symmetric nuclear matter at several times nuclear saturation density,
with classic flow constraints~\cite{Danielewicz2002} now complemented by
modern Bayesian and transport analyses~\cite{Sorensen2024PPNP,
OmanaKuttan2023PRL,Tsang2024NatAstron}.  Neutron-star observations probe
cold beta-equilibrated matter through massive pulsars~\cite{Antoniadis2013,
Cromartie2020,Romani2022}, the tidal deformability measured in
GW170817~\cite{Abbott2017PRL,Abbott2018PRL,Abbott2019PRX}, and NICER
mass-radius data~\cite{Riley2019,Miller2019,Riley2021,Miller2021,
Choudhury2024,Rutherford2024}.  Together with microscopic input from chiral
effective field theory and perturbative QCD~\cite{Drischler2021,
Drischler2022,Keller2023PRL,Komoltsev2022,Gorda2023JHEP,Gorda2023ApJ}, these
data have made Bayesian dense-matter inference a central tool
~\cite{Raaijmakers2021,Legred2021,Huth2022,Brandes2023,Essick2023,
Jiang2023,Pang2024}.

A useful way to compare these constraints is to work with the dimensionless
trace measure, hereafter called the trace anomaly following common EOS usage,
\begin{equation}
  \Delta=\frac{1}{3}-\frac{P}{\varepsilon},
  \label{eq:delta}
\end{equation}
and the sound speed,
\begin{equation}
  \cs^2=\frac{\dd P}{\dd\varepsilon}.
  \label{eq:cs2}
\end{equation}
Fujimoto, Fukushima, McLerran, and Praszalowicz showed that the sound speed
can be decomposed as~\cite{Fujimoto2022PRL}
\begin{equation}
  \cs^2=\frac{P}{\varepsilon}
  -\frac{\dd\Delta}{\dd\ln\varepsilon}
  =
  \left(\frac{1}{3}-\Delta\right)
  -\frac{\dd\Delta}{\dd\ln\varepsilon}.
  \label{eq:fujimoto}
\end{equation}
It is related to the QCD trace of the energy-momentum tensor by
\(\Theta^\mu_{\ \mu}=\varepsilon-3P=3\varepsilon\Delta\).  The derivative term
therefore connects the rise of \(\cs^2\) to the rate at which dense matter
approaches a more conformal thermodynamic trajectory.  Recently, Li used
collective-flow information to infer an exploratory trace reference for cold
dense matter from laboratory data~\cite{Li2026TraceAnomalyFlow}.  The
appearance of the same thermodynamic variables in HIC and neutron-star
inference suggests a sharper question than whether dense matter is stiff or
soft: how is the nonconformal trajectory of symmetric matter mapped into the
beta-equilibrated matter inside neutron stars?

This question is not answered by an EOS band alone.  Macroscopic bands tell us
which thermodynamic trajectories are allowed, but they do not identify which
microscopic channel is associated with the trajectory.  Relativistic mean-field (RMF)
and covariant density functional models provide a natural channel language:
the scalar attraction, vector repulsion, and isovector interaction determine
the saturation mechanism, high-density stiffness, and symmetry sector
~\cite{Serot1986,Ring1996,Vretenar2005,Niksic2011}.  Bayesian RMF studies
have already shown the importance of combining nuclear and
astrophysical information~\cite{Malik2022,Salinas2023}.  Here we use an
inverse-mapped RMF ensemble to turn the trace anomaly into a microscopic
diagnostic.  In contrast to a single-channel high-density deformation, the
inverse ensemble samples scalar, vector, and isovector density dependences
simultaneously.  The new point is therefore not another allowed EOS band, but
a test of which density-dependent RMF channel carries the trace response when
the HIC isoscalar trajectory is mapped into beta-equilibrated neutron-star
matter.

Our main result is that, within the present inverse-mapped DD-PC ensemble, the
trace measure carries a measurable isospin splitting.  The symmetric-matter
posterior captures the overall flow-based trace trend, while the
beta-equilibrated posterior approaches the neutron-star
trace bands more slowly.  This difference has a leading posterior correlation
with the density derivative of the isovector coupling at \(2--3\nzero\), and
with a mixture of isovector and scalar-vector rearrangement effects at higher
density.  Thus the trace anomaly is not only a possible signature of the
approach to conformality; in this inverse ensemble it is also a
composition-sensitive diagnostic of the high-density symmetry sector.  Its
usefulness is not that the static quantity
\(\Delta=1/3-P/\varepsilon\) contains information independent of
\(P/\varepsilon\), but that the SNM--beta splitting and the associated
derivative term separate the isoscalar trace trajectory from the isovector
composition map.

\section{Inverse-mapped RMF ensemble}

We use a zero-temperature DD-PC-like RMF model for uniform matter, with
density-dependent contact couplings in the scalar, isoscalar-vector, and
isovector-vector channels.  For nucleon density
\(n=n_n+n_p\), scalar density \(n_s\), and isovector density
\(n_3=n_p-n_n\), the interaction part of the energy density is
\begin{equation}
  \varepsilon_{\rm int}
  =
  \frac{1}{2}G_\omega(n)n^2
  +\frac{1}{2}G_\sigma(n)n_s^2
  +\frac{1}{2}G_\rho(n)n_3^2 ,
  \label{eq:energy-functional}
\end{equation}
with \(M^*=M-G_\sigma n_s\).  The density dependence is represented by a
Typel-Wolter-type rational form~\cite{Typel1999,Niksic2011},
\begin{equation}
  G_i(n)=G_{i0} f_i^2(n),\qquad
  f_i(n)=a_i\frac{1+b_i(x+d_i)^2}{1+c_i(x+d_i)^2},
  \quad x=\frac{n}{n_0},
  \label{eq:twmap}
\end{equation}
for \(i=\sigma,\omega,\rho\).  The normalization \(f_i(n_0)=1\) fixes the
couplings at saturation, while the asymptotic values
\(f_{\sigma,\infty}\), \(f_{\omega,\infty}\), and \(f_{\rho,\infty}\) are
sampled.  The inverse map follows the construction of
Ref.~\cite{XieXiaInverseMapping}; we summarize the ingredients needed here.
The saturation strengths are fixed sequentially: \(G_{\sigma0}\) follows from
the chosen Dirac effective mass, \(G_{\omega0}\) from the SNM binding energy
at \(n_0\), and \(G_{\rho0}\) from \(E_{\rm sym}(n_0)\).  The isovector
shape is then fixed by four constraints,
\(f_\rho(1)=1\), \(f_{\rho,\infty}\), and the first two derivatives at
sat\-uration, which are determined by \(L\) and \(K_{\rm sym}\).  For the
isoscalar shapes we use the same reduced Typel-Wolter branch as in
Ref.~\cite{XieXiaInverseMapping}: the asymptotic values and the normalization
are imposed analytically, and the remaining two shape parameters are obtained
from \(P(n_0)=0\) and the input incompressibility \(K_0\), with the
rearrangement term included.  Thus the ten sampled macroscopic quantities
define a determined DD-PC-like uniform-matter interaction rather than freely
fitted microscopic coefficients in Eq.~\eqref{eq:twmap}.

This ``inverse map'' should be understood in a restricted sense.  Within the
chosen rational ansatz and the reduced Typel-Wolter branch, each accepted
macroscopic parameter vector is mapped deterministically to a set of uniform
matter couplings.  We do not claim a globally unique inverse of the RMF
problem: other functional forms, additional density structures, or different
choices of branch could represent the same low-order nuclear-matter
parameters.  Samples that do not yield a physical branch, a stable EOS, or a
causal trace-valid trajectory in the density interval used below are rejected.
A full Jacobian analysis of possible multiple branches belongs to the
construction of a more complete inverse-EDF family; here the fixed branch is
used as a controlled channel basis in which to ask how trace diagnostics are
projected onto scalar, vector, and isovector directions.

The pressure includes the rearrangement contribution required by
thermodynamic consistency in density-dependent RMF models,
\begin{equation}
  \Sigma_R =
  \frac{1}{2}n^2\frac{\dd G_\omega}{\dd n}
  -\frac{1}{2}n_s^2\frac{\dd G_\sigma}{\dd n}
  +\frac{1}{2}n_3^2\frac{\dd G_\rho}{\dd n}.
  \label{eq:rearrangement}
\end{equation}
For SNM the pressure is computed from the analytic mean-field expression
containing \(n\Sigma_R\); as a thermodynamic cross-check we compare it with
\(P=n\,\dd\varepsilon/\dd n-\varepsilon\) obtained from the independently
computed energy density.  For beta-equilibrated matter, charge neutrality and
beta equilibrium are solved first and the pressure is then evaluated from the
same thermodynamic identity.  This procedure also checks the
Hugenholtz-Van Hove relation
\(\mu_B=(\varepsilon+P)/n=\dd\varepsilon/\dd n\) on the EOS grid before the
trace-anomaly derivatives are formed.  For each posterior sample we compute
both symmetric nuclear matter (SNM) and beta-equilibrated matter on the same
density grid.  For the 167 samples entering the continuous-curve bands, the
maximum relative HVH residual over \(1.2\le n/n_0\le6\) has a median
\(5.8\times10^{-6}\), a 95th percentile \(6.9\times10^{-6}\), and a maximum
\(7.5\times10^{-6}\).  The residual is even smaller at the densities used for
the correlation analysis, with median values decreasing from
\(3.6\times10^{-6}\) at \(2n_0\) to \(5.1\times10^{-7}\) at \(5n_0\).

The present ensemble is used as a uniform-matter inverse map.  The main
posterior used below combines chiral EFT, HIC flow, and neutron-star
information; additional posteriors with selected subsets of these inputs are
used only to diagnose data-combination dependence.  It is therefore not a
finite-nucleus-calibrated covariant density functional.  This is a deliberate
separation of questions: the present work asks which uniform-matter DD-PC-like
channels are associated with dense-matter trace diagnostics, not whether every
posterior sample is already a predictive functional for finite nuclei.
Finite-nucleus data constrain additional surface, gradient, shell, and pairing
sectors that are absent from the uniform-matter calculation.  They would
therefore act as an additional filter, or as an additional layer in a future
finite-nucleus-protected refit, rather than as a replacement for the present
trace diagnostic.  This distinction is important: the neutron-skin proxy
quoted below is only a low-density isovector sanity check and should not be
mistaken for a finite-nucleus fit.  The conclusions below should consequently
be read within
the uniform-matter inverse space defined by
Eqs.~\eqref{eq:energy-functional} and \eqref{eq:twmap}: they point to the
microscopic channel directions that a predictive finite-nucleus-calibrated
functional would have to preserve if it is to reproduce the same HIC--NS trace
mapping.

We nevertheless perform a low-density isovector sanity check.  The posterior
for the full data set gives \(L=44.1^{+19.4}_{-11.5}\) MeV and
\(K_{\rm sym}=-76.2^{+93.1}_{-54.9}\) MeV at 90\% credibility.  Using two
standard empirical EDF correlations between \(L\) and the neutron skin of
\(^{208}\mathrm{Pb}\), these values correspond to
\(R_{\rm skin}^{208}=0.156\)--\(0.166\) fm for the median estimate, with a
90\% range approximately \(0.139\)--\(0.194\) fm depending on the correlation used.
This is not a finite-nucleus calculation, but it indicates that the inferred
isovector sector is not obviously outside the empirical nuclear-structure
domain.
We also verified that the main trace diagnostic is insensitive to applying
this proxy as a posterior filter.  Requiring both empirical skin estimates to
lie in the conservative interval \(0.12\)--\(0.22\) fm removes only one of the
3187 trace-valid samples.  A tighter interval, \(0.14\)--\(0.20\) fm, keeps
3061 samples and leaves the key correlation essentially unchanged:
\(r_s(\Delta_{\SNM}-\Delta_{\betaeq},\dd f_\rho/\dd n)=0.927\) at
\(2n_0\) and \(0.897\) at \(3n_0\), compared with 0.931 and 0.900 before the
filter.  The probability that \(\Delta_{\SNM}-\Delta_{\betaeq}>0\) remains
unity from \(2n_0\) to \(5n_0\).  Thus the trace-isospin result does not
originate from the small part of the posterior that would be disfavored by
empirical low-density neutron-skin systematics.

\begin{table}[t]
\centering
\caption{Uniform priors used in the inverse-RMF posterior.  Energies are in
MeV and \(n_0\) is in \(\mathrm{fm}^{-3}\).}
\label{tab:priors}
\begin{tabular}{lll}
\toprule
Parameter & Meaning & Prior \\
\midrule
\(K_0\) & incompressibility & \(220\)--\(260\) \\
\(M^*/M\) & Dirac effective mass & \(0.45\)--\(0.65\) \\
\(n_0\) & saturation density & \(0.145\)--\(0.170\) \\
\(E_0\) & binding energy & \(-16.5\)--\(-15.8\) \\
\(E_{\rm sym}\) & symmetry energy & \(28.5\)--\(34.9\) \\
\(L\) & symmetry-energy slope & \(20\)--\(120\) \\
\(K_{\rm sym}\) & symmetry incompressibility & \(-400\)--\(100\) \\
\(f_{\sigma,\infty}\) & scalar high-density limit & \(0.3\)--\(0.9\) \\
\(f_{\omega,\infty}\) & vector high-density limit & \(0.3\)--\(1.4\) \\
\(f_{\rho,\infty}\) & isovector high-density limit & \(0.2\)--\(0.9\) \\
\bottomrule
\end{tabular}
\end{table}

The posterior is substantially narrower than the prior in the parameters most
relevant for the isovector conclusion.  The 90\% posterior widths of
\(L\), \(K_{\rm sym}\), and \(f_{\rho,\infty}\) are 0.309, 0.296, and 0.278
of their prior widths, respectively.  Thus the strong correlations reported
below are not simply a restatement of the original prior volume in the
symmetry sector.

The likelihood is a product of three data projections.  Chiral EFT enters as
a Gaussian penalty to the pressure band over \(0.08\le n\le0.16\)
\(\mathrm{fm}^{-3}\); HIC flow enters as a Gaussian penalty to the symmetric
matter pressure band over \(2\le n/n_0\le4.5\); and neutron-star information
enters through source-specific NICER mass-radius kernel-density estimates for
PSR J0030+0451, PSR J0437-4715, and PSR J0740+6620 evaluated along the TOV
sequence.  The optional massive-pulsar existence term is not used in the main
posterior to avoid double counting mass information already present in the
NICER posteriors.  The ten-dimensional posterior is sampled with
PyMultiNest/MultiNest using 2000 live points~\cite{Feroz2009MultiNest,
Buchner2014PyMultiNest}.
As a convergence sanity check, we compared the production chain with an
independent backup run using the same likelihood and prior.  The production
and backup runs contain 10991 and 10221 equal-weight posterior samples,
respectively, and give \(\ln Z=-15.716\pm0.082\) and
\(-15.462\pm0.082\).  The marginal posterior medians of all ten inverse-map
parameters agree within 0.14 of the production-run 68\% credible width; the
largest normalized shift occurs for \(L\).  The channel-correlation analysis
below is therefore not based on an under-sampled 10-dimensional posterior.
Perturbative-QCD calculations are used here only as high-density context in
the introduction, not as a likelihood term.  The reason is that our diagnostic
focus is the \(2--5n_0\) interval where HIC and neutron-star matter overlap
most directly; enforcing a pQCD matching prior would answer a different
question about asymptotic completion of the EOS.

For each sample we compute
\begin{equation}
  \Delta(n)=\frac{1}{3}-\frac{P(n)}{\varepsilon(n)},
  \qquad
  \cs^2(n)=\frac{\dd P}{\dd\varepsilon},
  \label{eq:observables}
\end{equation}
and the derivative contribution
\begin{equation}
  D(n)\equiv-\frac{\dd\Delta}{\dd\ln\varepsilon}.
  \label{eq:derivterm}
\end{equation}
The derivatives are evaluated on the EOS grid with a local linear slope in
\(\varepsilon\) or \(\ln\varepsilon\), using the same prescription for SNM
and beta-equilibrated matter.  The equality
\(\cs^2=P/\varepsilon+D\) is then used as a thermodynamic check, so the
sound speed entering the figures is tied to the trace-anomaly decomposition
rather than to an independent smoothing convention.
We also evaluate the composition splitting
\begin{equation}
  \delta_{\rm iso}X
  \equiv X_{\SNM}-X_{\betaeq},
  \qquad
  X\in\{\Delta,\cs^2,D,P/\varepsilon\},
  \label{eq:splitting}
\end{equation}
at fixed \(n/\nzero\) and at fixed \(\varepsilon/\epszero\).  The latter is
important because HIC and neutron-star constraints are often compared in the
\(\Delta(\varepsilon)\) plane rather than at the same baryon density.

The MultiNest run contains 10991 equal-weight posterior samples, obtained with
2000 live points and a nested-sampling evidence
\(\ln Z=-15.716\pm0.082\).  Reprocessing this full posterior gives 3187
samples that pass the discrete trace-stability analysis.  The continuous band
calculation is stricter because every sample must remain thermodynamically
stable on the full grid \(1\le n/\nzero\le 6\); 167 samples pass this
continuous-curve filter.  The plotted continuous bands use these valid
continuous curves, while the coupling-response correlations and
signal-to-width tests use the 3187 trace-valid samples from the full
posterior.  This
separation avoids discarding otherwise useful posterior samples from the
channel-correlation analysis merely because they fail the more stringent
continuous-curve requirement at the edge of the plotted density interval.
For correlation coefficients we quote bootstrap uncertainties from 2000
resamplings of the 3187 trace-valid samples.  The main isovector result is stable:
for \(\delta_{\rm iso}\Delta\) versus \(\dd f_\rho/\dd n\),
\(r_s=0.931^{+0.003}_{-0.003}\) at \(2n_0\) and
\(r_s=0.900^{+0.004}_{-0.004}\) at \(3n_0\).  The same bootstrap analysis is
used to check that this correlation is not a proxy for the beta-equilibrium
proton fraction.
For the data-combination test we repeat the same trace analysis for
posteriors constrained by neutron-star data only, HIC data only, chiral EFT
only, and the corresponding two-input combinations.  These auxiliary runs are
used to assess which data projection most strongly shapes the microscopic
correlations, not
to replace the full posterior used for the main quantitative statements.

\section{Thermodynamic comparison}

\begin{figure}[H]
  \centering
  \includegraphics[width=0.98\textwidth]{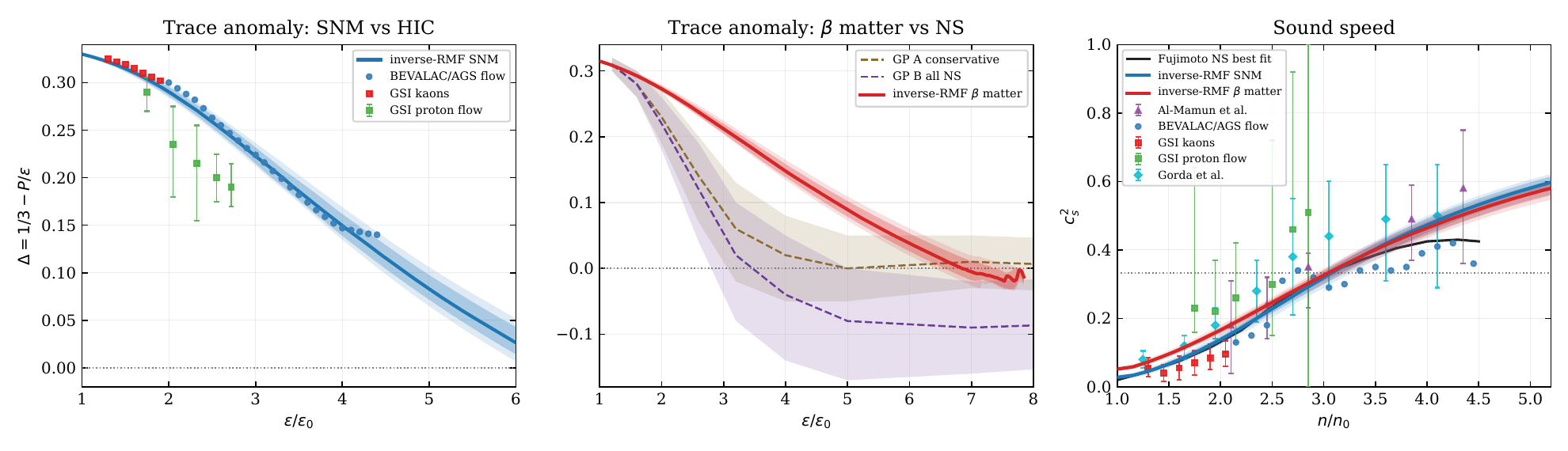}
  \caption{Inverse-RMF thermodynamic comparison.  The left panel compares the
  SNM trace measure with exploratory flow-based HIC points from
  Ref.~\cite{Li2026TraceAnomalyFlow}; the middle panel compares
  beta-equilibrated matter with the neutron-star GP bands of
  Ref.~\cite{Fujimoto2022PRL}; the right panel shows the sound-speed
  evolution and the conformal value \(1/3\).  Bands denote posterior
  intervals of the inverse-mapped RMF ensemble.}
  \label{fig:benchmark}
\end{figure}

Figure~\ref{fig:benchmark} compares the inverse-RMF posterior bands with the
digitized trace-anomaly and sound-speed reference results.  The left panel
shows SNM against the cold-SNM trace trajectory inferred from collective-flow
transport analyses~\cite{Li2026TraceAnomalyFlow}.  We use these points as
exploratory laboratory anchors rather than as direct \(T=0\) measurements:
thermal, nonequilibrium, and transport-model uncertainties are part of the
extraction systematics.  With this interpretation, the median inverse
trajectory passes through the BEVALAC/AGS and GSI kaon points, while
the proton-flow subset is followed less closely but still constrains the
overall downward trend.  Quantitatively, the root-mean-square distances to
the BEVALAC/AGS and GSI kaon trace points are \(0.0066\) and \(0.0044\),
respectively, while the proton-flow subset gives \(0.0489\).

The middle panel compares beta-equilibrated matter with the neutron-star GP
bands of Ref.~\cite{Fujimoto2022PRL}.  The beta-matter posterior decreases
more slowly than the GP medians.  We do not interpret this as an immediate
failure.  Instead, it identifies the central issue of the present work:
the HIC trace is primarily an isoscalar SNM constraint, whereas the
neutron-star trace includes beta equilibrium and therefore the high-density
isovector response.  The right panel shows that both SNM and beta matter
cross the conformal sound-speed value \(1/3\) near \(3\nzero\), with median
crossings at \(n/\nzero=3.079\) and \(3.042\), respectively.  At
\(5\nzero\), the medians give \(\Delta_{\SNM}=0.063\),
\(\Delta_{\betaeq}=0.054\), \(\cssnm^2=0.581\), and
\(\csbeta^2=0.564\).

\section{Isospin splitting of the trace measure}

\begin{figure}[H]
  \centering
  \includegraphics[width=0.98\textwidth]{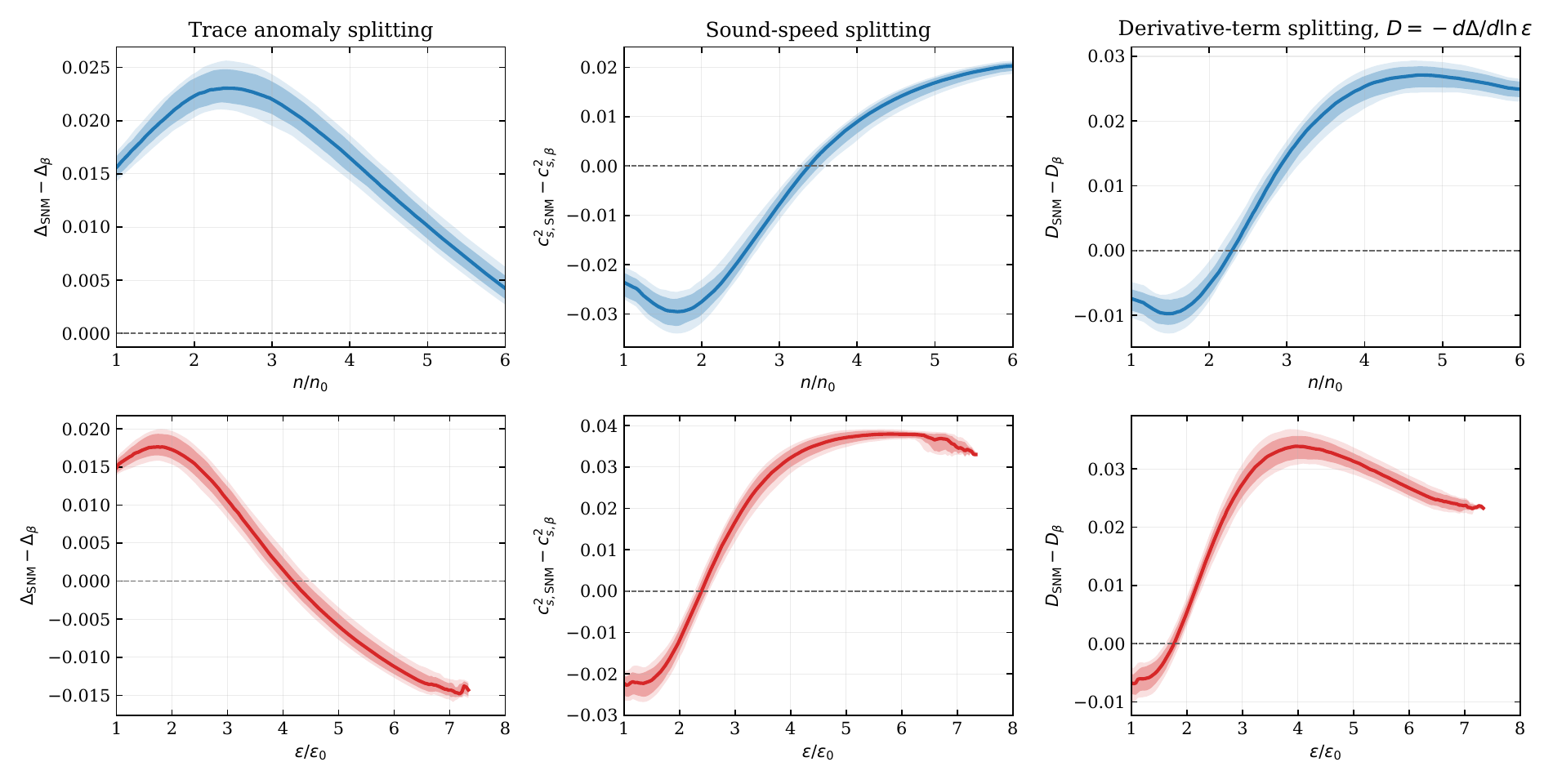}
  \caption{Isospin splitting of the inverse-RMF trace diagnostics.  The top
  row shows differences at fixed baryon density, and the bottom row shows
  differences at fixed energy density.  Shaded regions denote the 68\% and
  90\% posterior intervals; solid curves show the medians.}
  \label{fig:isospin}
\end{figure}

The thermodynamic comparison motivates a direct calculation of the SNM--beta
splitting.  Figure~\ref{fig:isospin} shows the median and posterior intervals
for \(\Delta_{\SNM}-\Delta_{\betaeq}\),
\(\cssnm^2-\csbeta^2\), and \(D_{\SNM}-D_{\betaeq}\).  At fixed
density, the trace-anomaly splitting is positive throughout
\(2--5\nzero\), decreasing from about \(0.022\) at \(3\nzero\) to
about \(0.010\) at \(5\nzero\).  The sound-speed splitting is more
structured: it is negative below \(3\nzero\), crosses zero near
\(3.38\nzero\), and becomes positive at higher density.  Thus beta matter is
slightly stiffer at lower density, while SNM becomes stiffer once the
derivative term has fully turned on.
The positive trace splitting is not a marginal posterior fluctuation.  Using
the 3187 trace-valid posterior samples, the probability
\(P(\Delta_{\SNM}-\Delta_{\betaeq}>0)\) is unity at all reference densities
from \(2\nzero\) to \(5\nzero\).  The posterior signal-to-width ratio
\[
  S(n)=
  \frac{\mathrm{median}(\Delta_{\SNM}-\Delta_{\betaeq})}
       {\sigma_{68}(\Delta_{\SNM}-\Delta_{\betaeq})}
\]
is \(14.7\), \(14.0\), \(11.4\), and \(7.9\) at
\(2\nzero\), \(3\nzero\), \(4\nzero\), and \(5\nzero\), respectively.

\begin{table}[!htbp]
\centering
\caption{Median SNM--beta splitting in the inverse-mapped RMF ensemble.  The
derivative term is \(D=-\dd\Delta/\dd\ln\varepsilon\).}
\label{tab:isospin}
\begin{tabular}{cccc}
\toprule
\(n/\nzero\) &
\(\Delta_{\SNM}-\Delta_{\betaeq}\) &
\(\cssnm^2-\csbeta^2\) &
\(D_{\SNM}-D_{\betaeq}\) \\
\midrule
2.0 & 0.0224 & -0.0276 & -0.0054 \\
2.8 & 0.0227 & -0.0120 &  0.0108 \\
3.0 & 0.0221 & -0.0076 &  0.0145 \\
4.0 & 0.0166 &  0.0090 &  0.0253 \\
5.0 & 0.0101 &  0.0169 &  0.0269 \\
\bottomrule
\end{tabular}
\end{table}

At fixed energy density the same physics appears in a different projection.
The median trace-anomaly splitting crosses zero at
\(\varepsilon/\epszero\simeq4.17\), while the sound-speed and derivative-term
splittings cross earlier, near \(2.40\) and \(1.77\), respectively.  The
projection dependence is expected because SNM and beta matter assign
different energy densities to the same baryon density once the isovector
sector and beta equilibrium are included.  It also provides a useful warning:
HIC and neutron-star trace bands should not be compared as if they were the
same composition.

\FloatBarrier

\section{Microscopic channel origin}

The next question is whether the splitting can be assigned to a microscopic
channel.  We correlate the trace diagnostics at fixed density with the
couplings and their density derivatives.  The strongest posterior
correlations are with the derivative of the isovector coupling.  For
\(\Delta_{\SNM}-\Delta_{\betaeq}\), the Spearman coefficients with
\(\dd f_\rho/\dd n\) are
\(r_s=0.931^{+0.003}_{-0.003}\),
\(0.911^{+0.003}_{-0.004}\), and
\(0.900^{+0.004}_{-0.004}\) at \(2.0\nzero\), \(2.8\nzero\), and
\(3.0\nzero\), respectively, where the errors are 68\% bootstrap intervals.
The same information appears with the opposite sign in the
\(P/\varepsilon\) splitting, as required by Eq.~\eqref{eq:delta}.  The
traditional low-density symmetry-sector parameters are also informative, but
less directly: \(L\) and \(K_{\rm sym}\) correlate with the trace splitting at
the \(r_s\simeq0.4--0.7\) level over \(2--5\nzero\).
These correlations should be read as posterior diagnostics rather than as a
full refit in which one channel is varied while all finite-nucleus constraints
are reoptimized.  They indicate which directions in the inverse-mapped
DD-PC-like space carry the trace response.  A controlled local perturbation of
the isovector channel is introduced below as a mechanism check.

This correlation is not simply a proxy for the beta-equilibrium proton
fraction.  The correlation of \(\Delta_{\SNM}-\Delta_{\betaeq}\) with
\(Y_p\) is much weaker:
\(r_s=-0.042^{+0.018}_{-0.018}\) at \(2\nzero\) and
\(r_s=0.202^{+0.017}_{-0.017}\) at \(3\nzero\).  Thus the trace splitting is
not determined just by how many protons are present in beta matter; within the
inverse map it is tied mainly to the density dependence of the isovector
interaction.  This is the practical reason for using the trace splitting as
an isospin diagnostic rather than only plotting \(Y_p\), \(L\), or
\(c_s^2\): it is a thermodynamic composition response whose strongest
low-density posterior driver is the high-density derivative of the isovector
coupling.

At higher density the picture becomes less purely isovector.  The
derivative-term splitting at \(4\nzero\) correlates with
\(\dd f_\sigma/\dd n\), with \(r_s=0.897^{+0.004}_{-0.004}\), and also with
\(\dd f_\rho/\dd n\), with
\(r_s=0.788^{+0.008}_{-0.008}\).  At \(5\nzero\),
\(\dd f_\rho/\dd n\), \(\dd f_\sigma/\dd n\), and
\(\dd f_\omega/\dd n\) all correlate strongly with the derivative-term
splitting.  This means that the HIC-to-neutron-star trace mapping is
isovector dominated near \(2--3\nzero\), but the high-density derivative
term links the symmetry sector to the same scalar-vector rearrangement
associated with the overall nonconformal stiffening.

\begin{figure}[H]
  \centering
  \includegraphics[width=0.72\textwidth]{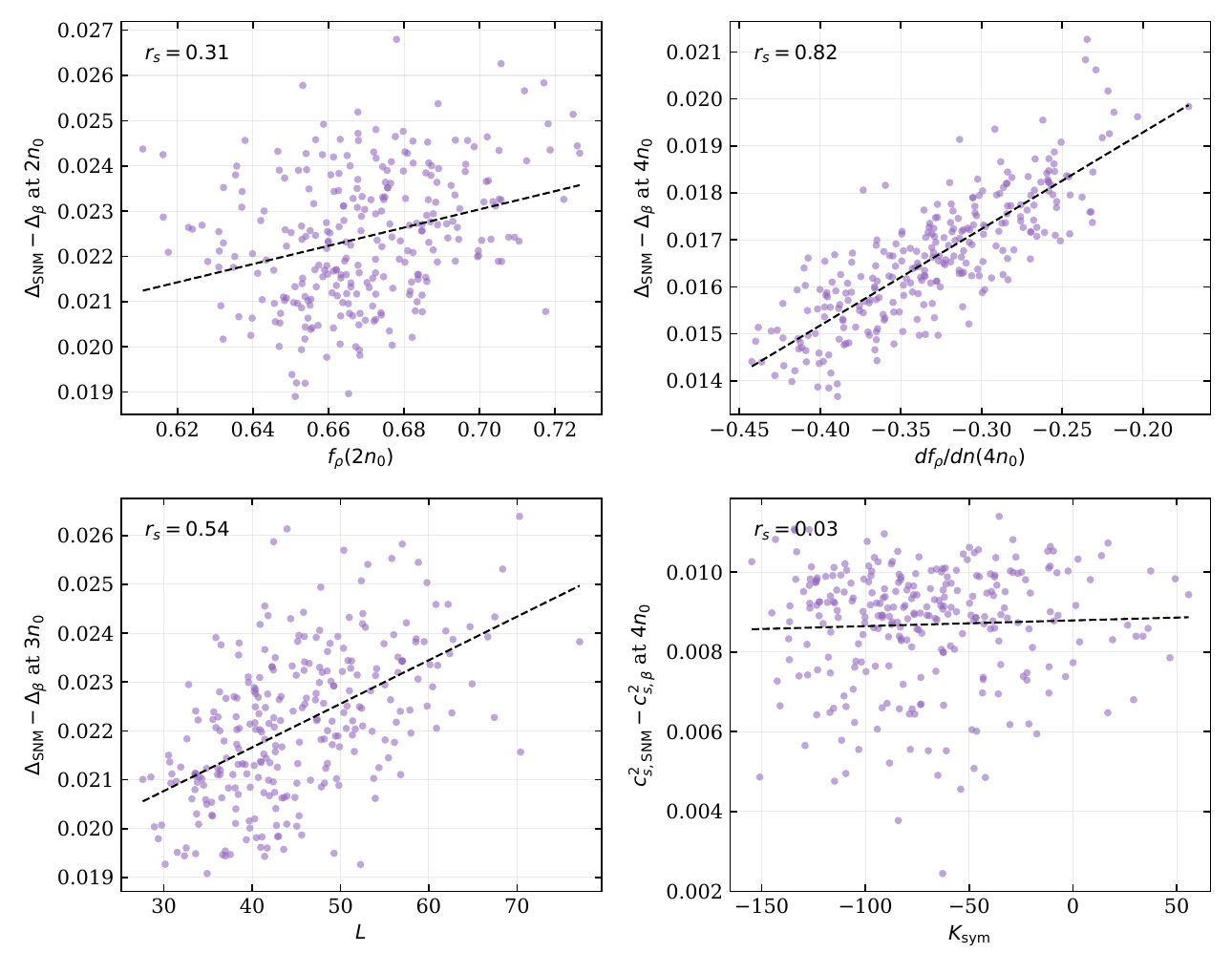}
  \caption{Representative microscopic correlations for the SNM--beta trace
  splitting.  The strongest low- and intermediate-density posterior correlate
  is the density derivative of the isovector coupling, while high-density
  sound-speed differences are less directly associated with \(K_{\rm sym}\).}
  \label{fig:correlations}
\end{figure}

\begin{figure}[H]
  \centering
  \includegraphics[width=0.72\textwidth]{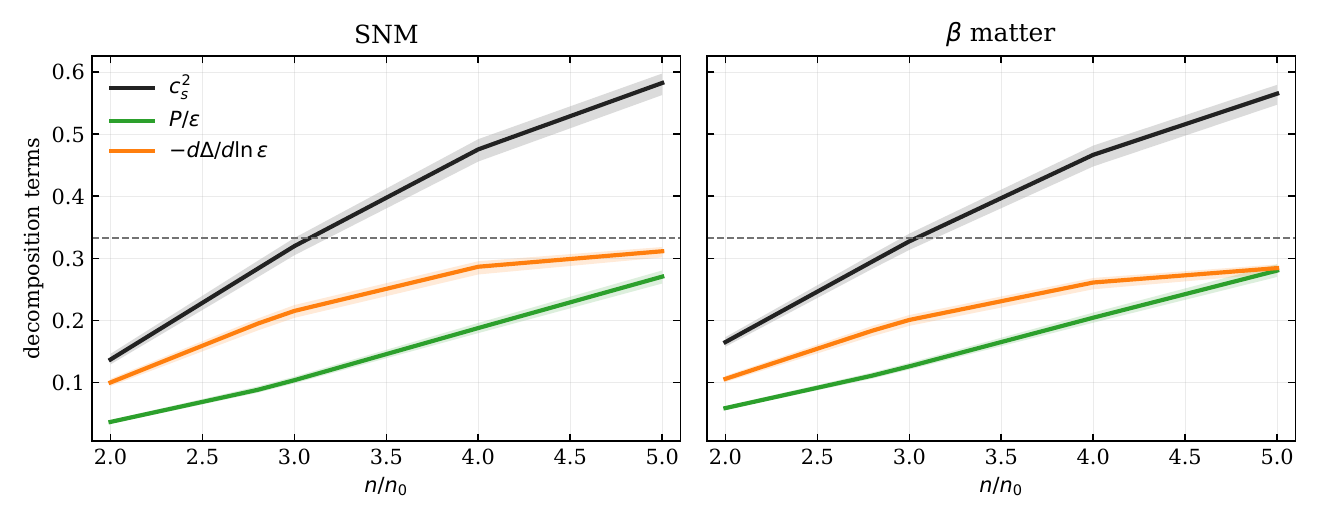}
  \caption{Fujimoto decomposition of the sound speed in the inverse-RMF
  ensemble.  The derivative term \(D=-\dd\Delta/\dd\ln\varepsilon\) provides
  an important contribution to the sound-speed increase.}
  \label{fig:decomp}
\end{figure}

Figure~\ref{fig:decomp} displays the Fujimoto decomposition for the inverse
ensemble.  The sound-speed rise is not produced solely by a large
\(P/\varepsilon\) term.  The derivative term \(D\) increases rapidly in the
same density interval in which \(\cs^2\) crosses \(1/3\).  A full
coupling-response scan gives two robust structures: the trace splitting at
\(2--3\nzero\) tracks the isovector derivative, while the high-density
derivative term is also associated with scalar-vector rearrangement.  We do
not interpret \(\Delta\) as statistically independent of the pressure ratio:
by definition, \(\delta_{\rm iso}\Delta=-\delta_{\rm iso}(P/\varepsilon)\).
The additional diagnostic value of the trace framework comes from keeping the
dimensionless pressure ratio and the slope term \(D\) in the same
conformality-based decomposition.

\begin{figure}[H]
  \centering
  \includegraphics[width=0.98\textwidth]{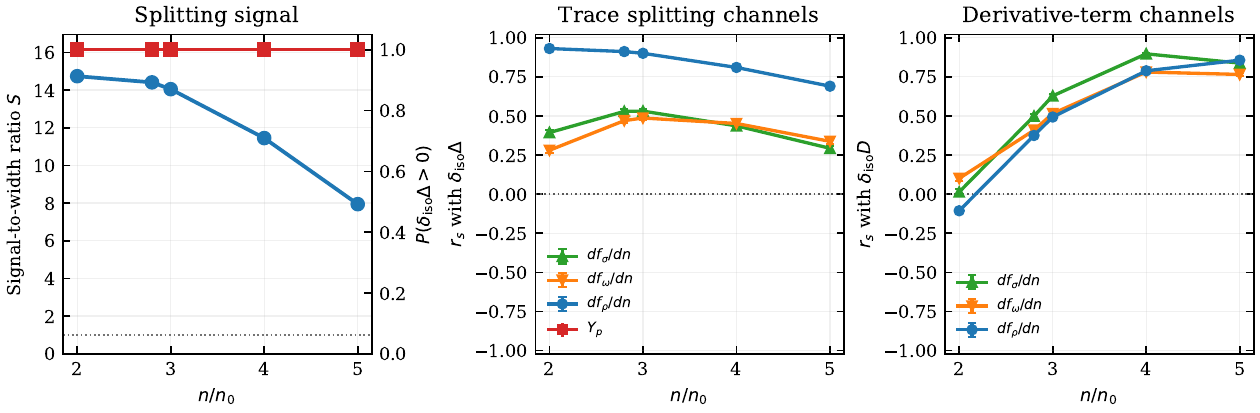}
  \caption{Signal-to-width ratio and channel association of the trace splitting.  Left:
  posterior signal-to-width ratio \(S=\mathrm{median}(\delta_{\rm iso}\Delta)/
  \sigma_{68}\) and the probability that \(\delta_{\rm iso}\Delta>0\).  Middle:
  signed channel-resolved Spearman correlations of
  \(\delta_{\rm iso}\Delta\) with \(\dd f_i/\dd n\) and with the
  beta-equilibrium proton fraction \(Y_p\).  Right: corresponding signed
  channel correlations for the derivative-term splitting
  \(\delta_{\rm iso}D\).  Error bars show 68\% bootstrap intervals.}
  \label{fig:sensitivity}
\end{figure}

\begin{table}[H]
\centering
\small
\caption{Posterior-median signed Spearman correlations used to compare the
diagnostic content of trace-related and conventional isovector quantities.
The first four rows use \(\dd f_\rho/\dd n\) as the driver.  The last three
rows correlate the trace splitting with the beta-equilibrium proton fraction
and with the low-density symmetry-sector parameters \(L\) and \(K_{\rm sym}\).
The first two rows have equal magnitude and opposite sign by
\(\Delta=1/3-P/\varepsilon\).}
\label{tab:observable-info}
\begin{tabular}{lcccc}
\toprule
Correlation pair & \(2n_0\) & \(3n_0\) & \(4n_0\) & \(5n_0\) \\
\midrule
\(\delta_{\rm iso}(P/\varepsilon)\) vs. \(\dd f_\rho/\dd n\) & \(-0.931\) & \(-0.900\) & \(-0.810\) & \(-0.690\) \\
\(\delta_{\rm iso}\Delta\) vs. \(\dd f_\rho/\dd n\) & \(+0.931\) & \(+0.900\) & \(+0.810\) & \(+0.690\) \\
\(\delta_{\rm iso}c_s^2\) vs. \(\dd f_\rho/\dd n\) & \(-0.610\) & \(-0.278\) & \(+0.109\) & \(+0.288\) \\
\(\delta_{\rm iso}D\) vs. \(\dd f_\rho/\dd n\) & \(-0.105\) & \(+0.493\) & \(+0.788\) & \(+0.856\) \\
\(\delta_{\rm iso}\Delta\) vs. \(Y_p^\beta\) & \(-0.041\) & \(+0.202\) & \(+0.333\) & \(+0.388\) \\
\(\delta_{\rm iso}\Delta\) vs. \(L\) & \(+0.607\) & \(+0.522\) & \(+0.467\) & \(+0.428\) \\
\(\delta_{\rm iso}\Delta\) vs. \(K_{\rm sym}\) & \(+0.621\) & \(+0.653\) & \(+0.599\) & \(+0.526\) \\
\bottomrule
\end{tabular}
\end{table}

\begin{figure}[H]
  \centering
  \includegraphics[width=0.98\textwidth]{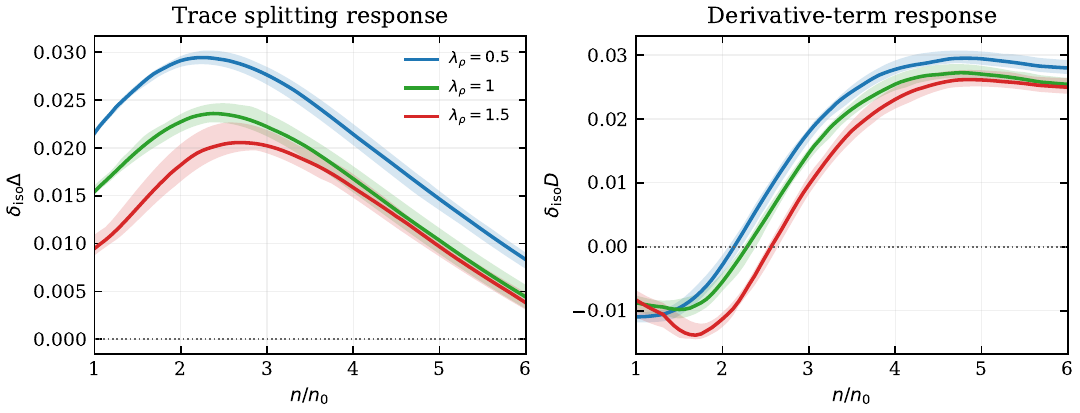}
  \caption{Controlled isovector-channel perturbation for 20 posterior samples
  with stable controlled curves.  The high-density departure of \(f_\rho(n)\)
  from its saturation value is scaled by \(\lambda_\rho\), while the scalar
  and isoscalar-vector channels are kept fixed.  Curves show medians and bands
  show 68\% intervals across the controlled sample ensemble.}
  \label{fig:control-rho}
\end{figure}

As a local mechanism check, we also perturb only the isovector coupling for
20 randomly drawn posterior samples that yield stable controlled curves.  We
replace
\[
  f_\rho(n) \rightarrow 1+\lambda_\rho[f_\rho(n)-1],
\]
which preserves the normalization \(f_\rho(n_0)=1\), and keep the scalar and
isoscalar-vector channels fixed.  Since \(n_3=0\) in SNM, this test primarily
changes beta-equilibrated matter and therefore isolates the local composition
map.  Figure~\ref{fig:control-rho} shows that increasing \(\lambda_\rho\)
from 0.5 to 1.5 systematically reduces the trace splitting near
\(2--4\nzero\) and shifts the derivative-term response across the ensemble.
At \(3n_0\), the median \(\Delta_{\SNM}-\Delta_{\betaeq}\) changes from
0.0276 to 0.0202 as \(\lambda_\rho\) increases from 0.5 to 1.5.  This does
not replace a global refit, but it shows that the leading posterior
correlation has the expected local isovector response.

\section{Discussion}

The analysis supports three conservative conclusions.  First, the flow-based
SNM trace anomaly and the neutron-star beta-matter trace anomaly are not
interchangeable constraints.  Their difference is small on the absolute
scale of \(\Delta\), but it is systematic and channel dependent.  Second, a
sound speed above \(1/3\) is not by itself a unique composition signature.
In the inverse-RMF ensemble it follows from the rapid decrease of
\(\Delta\), encoded in \(D=-\dd\Delta/\dd\ln\varepsilon\), and can be
produced within a hadronic mean-field channel description.  Third, the
composition mapping is a high-density symmetry-sector observable.  HIC flow
anchors mainly the isoscalar nonconformal path of SNM, while neutron-star
matter tests how this path changes under beta equilibrium.

These statements should not be overread.  The present calculation does not
rule out quark matter, hyperons, or other explicit degrees of freedom in
neutron-star cores.  It shows instead that the macroscopic trace-anomaly and
sound-speed patterns are not unique identifiers of those degrees of freedom.
Within the present inverse microscopic map, the same data also constrain
hadronic-coupling derivatives, especially the isovector response.
This provides a useful bridge between the trace-anomaly program of
Refs.~\cite{Fujimoto2022PRL,Li2026TraceAnomalyFlow} and ongoing efforts to
determine the high-density symmetry energy.

\begin{figure}[H]
  \centering
  \includegraphics[width=0.82\textwidth]{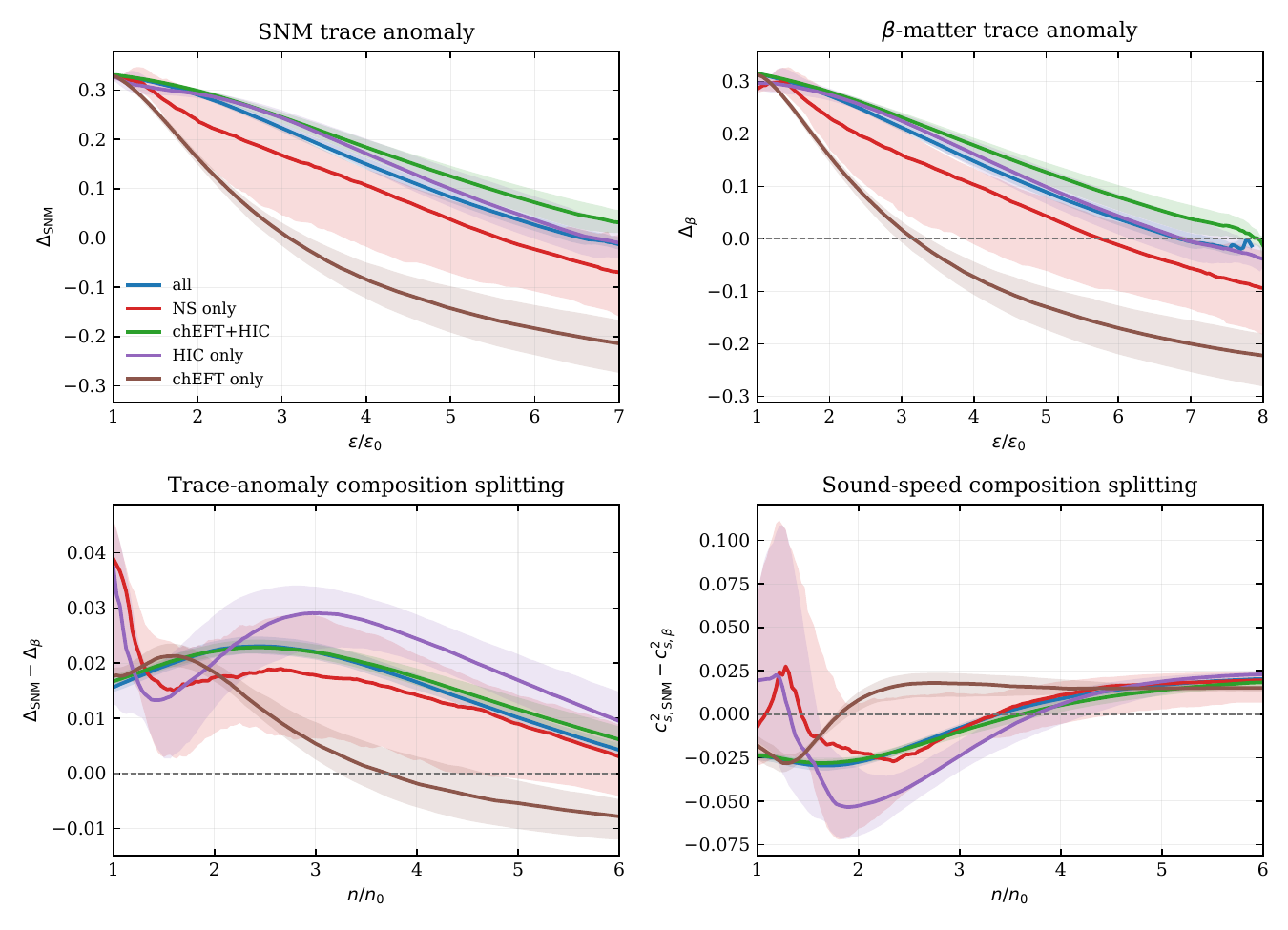}
  \caption{Dependence of the inverse-RMF trace diagnostics on the data
  combination.  The upper panels compare the SNM and beta-matter trace
  anomalies, while the lower panels show the corresponding composition
  splittings.  HIC information anchors the SNM trace trajectory, while the
  HIC+NS and full data combinations most clearly reveal the isovector
  association of \(\Delta_{\SNM}-\Delta_{\betaeq}\).  Neutron-star-only and
  \(\chi\)EFT-only combinations leave complementary directions weakly
  constrained.}
  \label{fig:data-combinations}
\end{figure}

The data-combination comparison in Fig.~\ref{fig:data-combinations} clarifies
why the combined analysis is needed.  Neutron-star data alone leave the SNM
trace band wide and make the strongest splitting correlations mostly
isoscalar; among the top 20 absolute Spearman correlations, only two are
isovector.  \(\chi\)EFT plus HIC data anchor the SNM trace but do not tightly
fix beta matter.  HIC-only and HIC+NS combinations, in contrast, make the
low- and intermediate-density splitting most strongly associated with
\(df_\rho/dn\), with 17 and 20 isovector entries, respectively, among the top
20 correlations.  The full posterior inherits this isovector separation while
retaining the neutron-star constraint on beta-equilibrated matter.

This behavior gives a simple physical interpretation of the data
complementarity.  Laboratory flow information primarily constrains the
isoscalar nonconformal trajectory of SNM.  Once that trajectory is anchored,
the residual uncertainty in translating SNM into beta-equilibrated matter is
projected mainly onto the symmetry sector.  Consequently, the trace anomaly
becomes a sensitive discriminator of the high-density isovector response only
when laboratory and astrophysical projections are combined.  The comparison
with conventional density-dependent RMF functionals below tests how much of
this association is generic to DD-RMF models and how much is preferred by the
inverse mapping.

The corresponding nested-sampling evidences are listed in
Table~\ref{tab:evidence}.  Because each row uses a different likelihood, these
numbers should not be read as model-selection Bayes factors between different
data sets.  They are instead a useful run diagnostic and a compact measure of
data-set compatibility within the same inverse-RMF prior.  The compatibility
index
\[
  \mathcal{I}_{A,B}=\ln Z_{A+B}-\ln Z_A-\ln Z_B
\]
is positive for HIC+NS and for the full combination, indicating that the HIC
and neutron-star projections select mutually compatible regions of the
inverse parameter space in the no-\(M_{\max}\) posterior used here.  The
negative values involving \(\chi\)EFT reflect the fact that low-density EFT
constraints carve away a different part of the prior volume; this is expected
because \(\chi\)EFT primarily fixes the saturation-neighborhood continuation,
whereas HIC and NS data shape the several-\(n_0\) trace trajectory.

\begin{table}[H]
\centering
\caption{Nested-sampling evidences for the no-\(M_{\max}\) inverse-RMF runs.
The compatibility index is defined only for combined data sets and is used as
a diagnostic, not as a model-ranking statistic between different likelihoods.}
\label{tab:evidence}
\begin{tabular}{lcc}
\toprule
Data set & \(\ln Z\) & Compatibility index \\
\midrule
NS & \(-2.659\pm0.037\) & -- \\
\(\chi\)EFT & \(-7.925\pm0.054\) & -- \\
HIC & \(-6.713\pm0.050\) & -- \\
\(\chi\)EFT+HIC & \(-15.436\pm0.080\) & \(-0.798\) \\
\(\chi\)EFT+NS & \(-11.787\pm0.057\) & \(-1.203\) \\
HIC+NS & \(-6.807\pm0.054\) & \(2.565\) \\
All & \(-15.716\pm0.082\) & \(1.580\) \\
\bottomrule
\end{tabular}
\end{table}

\begin{figure}[H]
  \centering
  \includegraphics[width=0.98\textwidth]{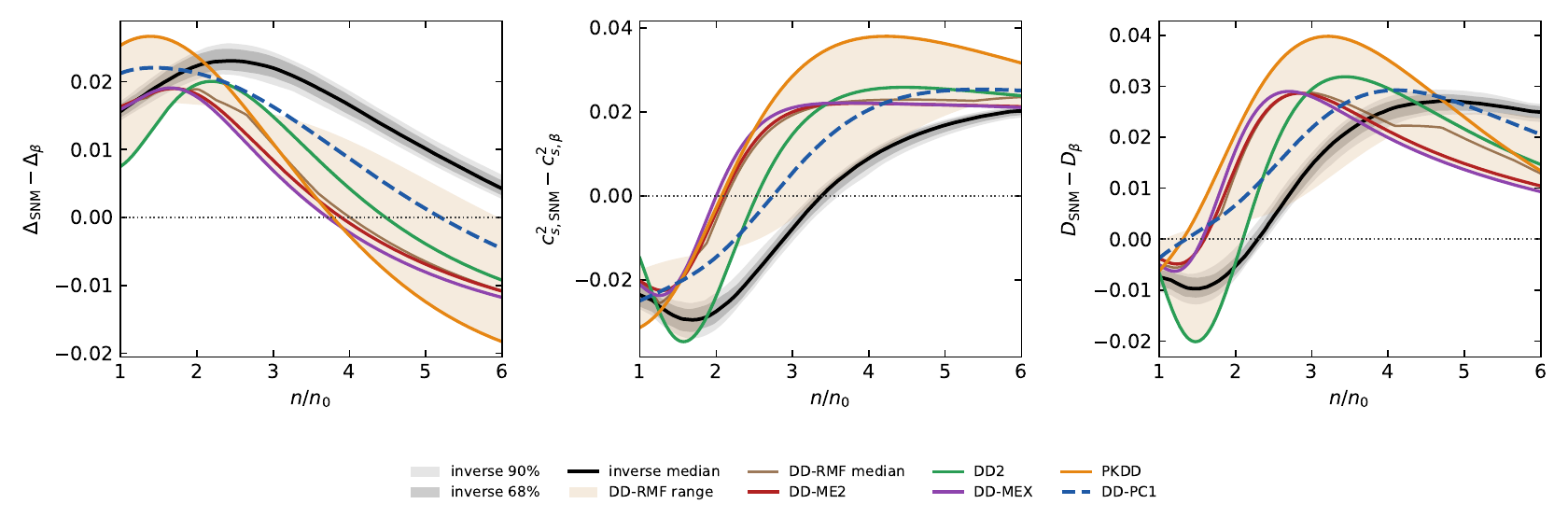}
  \caption{Comparison with conventional CDF baselines.  The gray bands show
  the inverse-RMF posterior intervals for the SNM--beta splittings.  The
  orange band and brown line show the range and median of the selected
  density-dependent fixed-RMF subset (TW99, DD-ME2, PKDD, DD2, DD-MEX, and
  DD-LZ1), with DD-ME2, DD2, DD-MEX, and PKDD highlighted.  DD-PC1 is shown
  as a point-coupling reference.  The inverse posterior overlaps parts of the
  standard DD-RMF landscape near \(2n_0\), but favors a more persistent
  high-density trace splitting.}
  \label{fig:cdf-baselines}
\end{figure}

As a baseline check against conventional covariant density functionals,
Fig.~\ref{fig:cdf-baselines} compares the inverse isospin-splitting band with
a fixed density-dependent RMF subset generated from the same uniform-matter
trace pipeline.  The fixed-RMF curves are not taken from mass--radius tables:
we recompute the zero-temperature SNM and beta-equilibrated EOS from the
tabulated RMF couplings, including the rearrangement term in the chemical
potentials and pressure.  The scalar-field equation for \(M^\ast\) is solved
with a bracketed root finder at each density, which avoids the high-density
fixed-point convergence failures that can otherwise masquerade as a physical
termination of the EOS.  The subset includes TW99~\cite{Typel1999},
DD-ME2~\cite{Lalazissis2005}, PKDD~\cite{Long2004PKDD},
DD2~\cite{Typel2010DD2}, DD-MEX~\cite{Taninah2020DDMEX}, and
DD-LZ1~\cite{Wei2020DDLZ1}; DD-PC1~\cite{Niksic2008} is kept as a
point-coupling reference.  We use the density-dependent subset for the
envelope because it is the closest conventional analogue of the inverse-mapped
density-dependent couplings; nonlinear constant-coupling RMF parameter sets
are not used to define this envelope.  The comparison is not a new calibration of these
functionals, but a diagnostic projection of their SNM--beta trace splitting.
This comparison shows that the inverse posterior is not isolated from standard DD-RMF
systematics: near \(2n_0\), PKDD and DD-PC1 lie inside the inverse 90\%
\(\Delta_{\SNM}-\Delta_{\betaeq}\) interval, and DD2 gives a compatible
sound-speed splitting.  At the same time, the full density dependence is more
selective.  At \(3n_0\), the inverse posterior gives
\(\Delta_{\SNM}-\Delta_{\betaeq}=0.0220\) with a 90\% interval
\(0.0195\text{--}0.0243\), while DD2 gives \(0.0148\), DD-ME2 gives
\(0.0084\), DD-MEX gives \(0.0069\), PKDD gives \(0.0109\), and DD-PC1 gives
\(0.0163\).  By \(5n_0\), the inverse median remains positive at \(0.0101\),
whereas DD-ME2, DD2, DD-MEX, DD-LZ1, and PKDD have crossed to negative
trace splitting, and DD-PC1 is close to zero.  Thus the
inverse-map result is not merely a consequence of using an RMF channel
language.  Rather, the data-favored inverse ensemble prefers a particular
persistent SNM--beta trace splitting within the broader DD-RMF landscape.

The main limitation is that the inverse ensemble used here is a uniform-matter
ensemble.  A predictive covariant density functional must also preserve
finite-nucleus ground-state properties.  This limitation does not invalidate
the internal channel diagnostic, because the diagnostic is formulated within a
fixed uniform-matter inverse space and compares SNM and beta matter generated
by the same microscopic couplings.  It does, however, limit the interpretation
of the posterior as a complete EDF.  The natural next step is therefore a
finite-nucleus filter or a finite-nucleus-protected functional family.  The
present result points to which microscopic directions such a refit should
test and, if stable, retain: the isovector derivative is the leading
posterior correlate of the SNM--beta trace splitting, while scalar-vector
rearrangement is linked to the derivative term associated with the
sound-speed peak.

\FloatBarrier

\section{Conclusion}

We have used an inverse-mapped RMF ensemble to ask whether trace anomaly and
sound-speed constraints can diagnose microscopic channel structure in the
composition map from HIC to neutron-star matter.  The ensemble follows the
flow-based SNM trace trend, whereas beta matter approaches the neutron-star
trace bands more slowly.  The resulting SNM--beta splitting is a robust
posterior-level
composition effect within this inverse ensemble.  At fixed density,
\(\Delta_{\SNM}-\Delta_{\betaeq}\) remains positive over \(2--5\nzero\),
while \(\cssnm^2-\csbeta^2\) changes sign near \(3.38\nzero\).
The splitting has its strongest posterior correlation with \(\dd f_\rho/\dd n\) at
\(2--3\nzero\), with scalar-vector derivatives becoming important correlates
of the high-density derivative term.

The trace anomaly is therefore more than an EOS-stiffness diagnostic.  Combined
with an inverse RMF map, it separates the isoscalar approach to conformality
from the isovector composition mapping between laboratory matter and
neutron-star matter.  The data-combination and controlled-\(f_\rho\) tests
indicate that this separation is well resolved within the posterior width and has
the expected local isovector response within the present inverse-mapped DD-PC
space.  A
stringent next test is to embed these channel directions in a
finite-nucleus-filtered or finite-nucleus-calibrated CDF family and determine
whether the same trace-anomaly isospin diagnostic survives when ground-state
nuclear data are imposed explicitly.  Although the present work does not
extract a model-independent band for \(E_{\rm sym}(n)\), it identifies the
SNM--beta trace splitting as a thermodynamic projection of the high-density
symmetry sector.  A natural follow-up is therefore to include this
trace-splitting information as an additional high-density symmetry-energy
constraint in a finite-nucleus-filtered inverse CDF analysis.

\bibliographystyle{elsarticle-num}
\bibliography{references}

\end{document}